\documentclass[journal,twoside,web]{ieeecolor}
\usepackage{tmi}
\usepackage{cite}
\usepackage{amsmath,amssymb,amsfonts}
\usepackage{algorithmic}
\usepackage{graphicx}
\usepackage{textcomp}
\def\BibTeX{{\rm B\kern-.05em{\sc i\kern-.025em b}\kern-.08em
    T\kern-.1667em\lower.7ex\hbox{E}\kern-.125emX}}
\usepackage{subcaption, multirow, array}
\usepackage{siunitx}

\usepackage[font=small,skip=5pt]{caption}
\usepackage{diagbox}
\usepackage{float}
\usepackage{hyperref}
\newcommand{\specialcell}[2][c]{%
  \begin{tabular}[#1]{@{}c@{}}#2\end{tabular}}
\newcolumntype{L}[1]{>{\raggedright\let\newline\\\arraybackslash\hspace{0pt}}m{#1}}
\newcolumntype{C}[1]{>{\centering\let\newline\\\arraybackslash\hspace{0pt}}m{#1}}
\newcolumntype{R}[1]{>{\raggedleft\let\newline\\\arraybackslash\hspace{0pt}}m{#1}}
\newcolumntype{P}[1]{>{\raggedright\arraybackslash}p{#1}}

\usepackage{xspace}
\makeatletter
\DeclareRobustCommand\onedot{\futurelet\@let@token\@onedot}
\def\@onedot{\ifx\@let@token.\else.\null\fi\xspace}

\def\etal{\emph{et al}\onedot}
\makeatother

\begin{document}

\title{autoTICI: Automatic Brain Tissue Reperfusion Scoring on 2D DSA Images of Acute Ischemic Stroke Patients}

\author{Ruisheng Su, Sandra A.P. Cornelissen, Matthijs van der Sluijs, Adriaan C.G.M. van Es, Wim H. van Zwam, Diederik W.J. Dippel, Geert Lycklama, Pieter Jan van Doormaal, Wiro J. Niessen, Aad van der Lugt, and Theo van Walsum

\thanks{Manuscript received January 15, 2021; revised March 22, 2021; accepted April 28, 2021. This work was supported by Health-Holland (TKI Life Sciences and Health)through the Q-Maestro project under Grant EMCLSH19006 and Philips Healthcare (Best, The Netherlands).}
\thanks{R. Su and T. van Walsum are with the Biomedical Imaging Group Rotterdam, Department of Radiology~\&~Nuclear Medicine, Erasmus MC, University Medical Center Rotterdam, The Netherlands~(e-mail:~r.su@erasmusmc.nl).}
\thanks{S.A.P. Cornelissen, M. van der Sluijs, P-J van Doormaal and A. van der Lugt are with the Department of Radiology~\&~Nuclear Medicine, Erasmus MC, University Medical Center Rotterdam, The Netherlands.}
\thanks{A.C.G.M. van Es is with the Department of Radiology, Leiden UMC, Leiden, The Netherlands.}
\thanks{W.H. van Zwam is with the Department of Radiology~\&~Nuclear Medicine, Maastricht UMC, Cardiovascular Research Institute Maastricht, The Netherlands.}
\thanks{D.W.J. Dippel is with the Department of Neurology, Erasmus MC, University Medical Center Rotterdam, The Netherlands.}
\thanks{G. Lycklama is with the Department of Radiology, Haaglanden Medical Center, The Hague, The Netherlands.}
\thanks{W.J. Niessen is with the Biomedical Imaging Group Rotterdam, Department of Radiology~\&~Nuclear Medicine, Erasmus MC, University Medical Center Rotterdam, The Netherlands, and with the Faculty of Applied Sciences, Delft University of Technology, The Netherlands.}
\thanks{Copyright (c) 2021 IEEE. Personal use is permitted. For any other purposes, permission must be obtained from the IEEE by emailing pubs-permissions@ieee.org. Digital Object Identifier 10.1109/TMI.2021.3077113}
}

\maketitle

\begin{abstract}
The Thrombolysis in Cerebral Infarction (TICI) score is an important metric for reperfusion therapy assessment in acute ischemic stroke. It is commonly used as a technical outcome measure after endovascular treatment (EVT). Existing TICI scores are defined in coarse ordinal grades based on visual inspection, leading to inter- and intra-observer variation. In this work, we present autoTICI, an automatic and quantitative TICI scoring method. First, each digital subtraction angiography (DSA) acquisition is separated into four phases (non-contrast, arterial, parenchymal and venous phase) using a multi-path convolutional neural network (CNN), which exploits spatio-temporal features. The network also incorporates sequence level label dependencies in the form of a state-transition matrix. Next, a minimum intensity map (MINIP) is computed using the motion corrected arterial and parenchymal frames. On the MINIP image, vessel, perfusion and background pixels are segmented. Finally, we quantify the autoTICI score as the ratio of reperfused pixels after EVT. On a routinely acquired multi-center dataset, the proposed autoTICI shows good correlation with the extended TICI (eTICI) reference with an average area under the curve (AUC) score of 0.81. The AUC score is 0.90 with respect to the dichotomized eTICI. In terms of clinical outcome prediction, we demonstrate that autoTICI is overall comparable to eTICI.
\end{abstract}


\section{Introduction}
\subsection{Clinical Background}

\IEEEPARstart{S}{troke} remains one of the worldwide leading causes of death and serious long-term disability~\cite{who2018global}. Due to the ageing population, stroke incidence is rising, which is posing a large and still increasing public health burden to the society. Ischemic stroke, which is caused by an occluded artery of the brain, is the most common stroke type, accounting for about 88\% of all strokes~\cite{higashida2003trial}.\par

Recent studies have shown that endovascular therapy (EVT) improves outcome in patients with acute ischemic stroke caused by a large vessel occlusion (LVO) in the anterior circulation~\cite{evt_prof, berkhemer2015randomized}. This led to novel studies on maximizing patient benefits before and during EVT. Many studies have focused on pre-interventional patient selection and outcome prediction based on scores such as Alberta Stroke Program Early CT Score (ASPECTS)~\cite{haussen2016automated} and collateral score~\cite{venema2017selection}. A recent study~\cite{dargazanli2017impact} demonstrated that peri-procedural imaging data can be valuable for predicting treatment outcome, which can in turn assist interventionalists to achieve better treatment quality.\par

DSA is the imaging modality to guide the EVT procedure. Despite rapid research progress in computational stroke related biomarker extraction from computed tomographic angiography (CTA)~\cite{su2020automatic} and CT Perfusion (CTP)~\cite{robben2020prediction} images, automatic peri-procedural imaging biomarker extraction from DSA is yet to be further explored.\par

One of the most widely adopted metrics for evaluation of EVT quality and prediction of functional outcome on DSA images is the so-called Thrombolysis In Cerebral Infarction (TICI) score~\cite{higashida2003trial} and its variants, such as modified TICI (mTICI)~\cite{ims2007interventional}, extended TICI (eTICI)~\cite{almekhlafi2014not} and expanded TICI~\cite{liebeskind2019etici}. The TICI scores define the extent of brain reperfusion. For example, the eTICI score is defined as follows: 

\begin{itemize}
    \item \textbf{eTICI 0}: no perfusion or antegrade flow in the target downstream territory (TDT). TDT refers to the occluded brain region that was supplied via antegrade blood flow prior to stroke onset \cite{zaidat2013recommendations}; 
    \item \textbf{eTICI 1}: blood flow past initial site of occlusion, but with minimal brain tissue perfusion;
    \item \textbf{eTICI 2A}: perfusion of $\leq 50\%$ of TDT;
    \item \textbf{eTICI 2B}: perfusion of $\geq 50\%$ of TDT;
    \item \textbf{eTICI 2C}: nearly complete perfusion except for slow flow or presence of small emboli in distal cortical vessels;
    \item \textbf{eTICI 3}: complete perfusion.
\end{itemize}

Above TICI score variants were introduced as attempts towards standardizing EVT treatment success scoring. However, these grading metrics suffer from several shortcomings:
\begin{itemize}
    \item Inter- and intra-observer variation: TICI assessment by visual inspection is inherently error prone, as it is subject to various factors, such as experience level of the rater and inspection attentiveness. Moreover, a recent study showed that TICI scores are generally overestimated by operators during EVT compared to core-lab raters~\cite{zhang2018operator};
    \item Coarse ordinal scale: TICI-like scores are mostly defined in 4-6 grades. Several studies have indicated that EVT outcome is associated with greater degrees of reperfusion~\cite{lansberg2012mri, khatri2005revascularization, tomsick2008revascularization, yoo2013refining, suh2013clarifying}. More granular perfusion grading systems would probably help to better assess treatment success;
    \item Conceptual confusion: in EVT terms, reperfusion is the antegrade restoration of a capillary blush, whereas recanalization generally refers to the restoration of blood flow past the arterial occlusion~\cite{zaidat2013recommendations}. In other words, recanalization is a necessary but insufficient condition for reperfusion. Although TICI scores define reperfusion scales, the concept of recanalization and reperfusion is often interchanged during visual TICI assessment in clinical practice~\cite{zaidat2013recommendations}. As a result, the cases of successful recanalization without adequate brain tissue reperfusion are usually overlooked, leading to overestimated TICI. 
\end{itemize}

To mitigate the above limitations of existing TICI scoring mechanisms, we pursue an objective, robust, and quantitative TICI scoring algorithm in a fully automated manner.

\subsection{Related Work}
In recent years, the field of computer vision evolved rapidly. Triggered by the advances in deep learning particularly, automated medical imaging analysis has gained momentum as well. 
In the field of stroke imaging, various algorithms on automatic quantification of imaging biomarkers have been proposed lately. These algorithms can be categorized into two groups based on the target imaging modality: pre-treatment imaging and peri-procedural imaging.\par

For pre-treatment prognosis, MRI, CTA and CTP images are widely used. Robben~\etal~\cite{robben2020prediction} exploited a multi-path CNN to predict the final infarct volume from CTP images and treatment metadata. Nielsen~\etal~\cite{nielsen2018prediction} demonstrated state-of-the-art performance of a deep CNN on final lesion volume and treatment outcome prediction on MRI images. Su~\etal~\cite{su2020automatic} utilized a U-Net architecture to quantify collateral scores from 3D CTA images. Related to brain perfusion estimation, McKinley~\etal~\cite{mckinley2017fully} developed an automated pipeline for penumbra volume estimation based on random forests using multi-modal MRI.\par

For peri-procedural evaluation of treatment effect, DSA is still considered the standard modality. Automatically quantified biomarkers on DSA have received less attention. As an example, Liebeskind~\etal~\cite{liebeskind2019automatic} recently proposed a machine learning based automatic arterial input function (AIF) extraction method. Concerning TICI scoring on DSA, to the best of our knowledge, no fully automatic algorithm has been proposed yet. Nevertheless, Prasetya~\etal~\cite{prasetya2020qtici} recently developed a quantitative TICI (qTICI) scoring pipeline which demands manual annotation. First, given a DSA acquisition under assessment, venous phase frames need to be manually removed by a neuro/interventional radiologist so to exclude the retrograde perfusion (via pial collaterals), as only antegrade (via recanalization) perfusion defines the reperfusion. Second, an estimated TDT is delineated by an expert observer for each artery occlusion of the DSA under assessment. Lastly, qTICI is defined as the percentage of reperfused pixels within TDT based on a threshold relative to the maximum intensity. Prasetya~\etal~\cite{prasetya2020qtici} demonstrated reliable correlation between qTICI and eTICI, revealing the potential value of quantitative TICI as an objective biomarker. Nevertheless, as qTICI relies on a series of manual steps, such a semi-automatic approach is more time consuming than traditional TICI grading procedures in practice. In addition, these manual steps are inevitably subjective, which can introduce new factors to inter- and intra-observer variability.\par

Beyond qTICI, we seek a fully automatic, yet explainable step by step approach for quantitative TICI scoring. \par

 \begin{figure*}[ht]
     \centering
        \includegraphics[clip, trim=0cm 3cm 0cm 1.9cm, width=\textwidth]{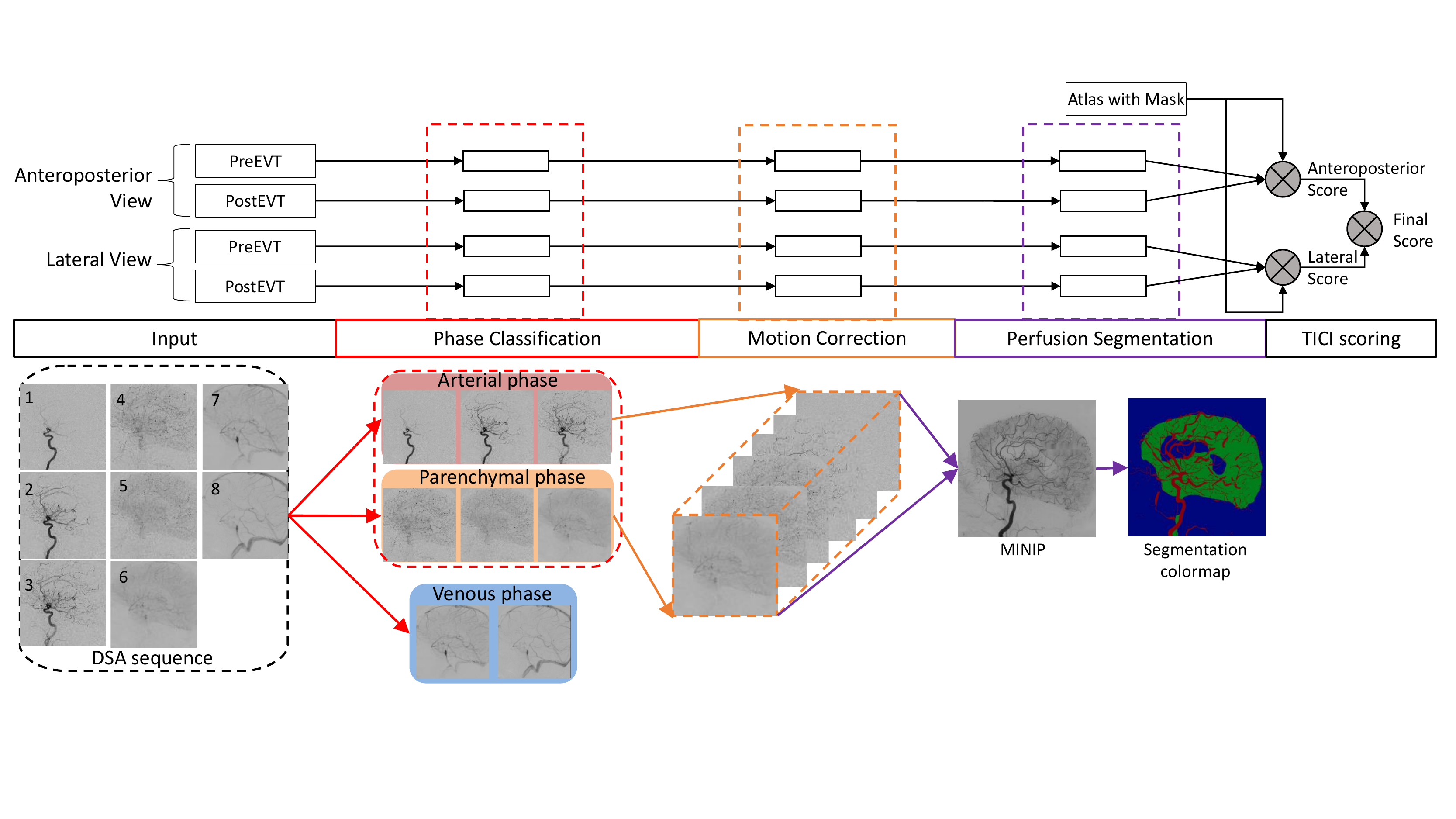}
        \caption{Proposed pipeline for automatic TICI scoring.}
        \label{fig:autotici_pipeline}
\end{figure*}

\subsection{Contributions}
The main contribution of this work is two-fold:
\begin{itemize}
    \item we propose a fully automatic and quantitative TICI scoring algorithm, as an objective and robust alternative to existing visual inspection based TICI scoring;
    \item we assess the proposed methods on a large multi-center data set acquired in clinical routine.
\end{itemize}

The remainder of this paper is organized as follows: Section~\ref{sec:method} provides a detailed overview of our proposed methods and Section~\ref{sec:data} describes the data used for experiments. Section~\ref{sec:exp&result} presents the experiment results and performance evaluations of the proposed methods, followed by further discussions in Section~\ref{sec:discussion}. Finally, Section~\ref{sec:conclusion} concludes the paper.\par

\section{Method} \label{sec:method}
Existing TICI scores are visually graded by estimating the extent of brain tissue-level reperfusion (post-EVT, numerator) in the initial TDT (pre-EVT, denominator). This would require a visual comparison between pre- and post-EVT acquisitions. Following this definition and the visual TICI scoring principles, we propose to automatically extract the numerator and denominator for quantitative TICI scoring.\par

As illustrated in Fig.~\ref{fig:autotici_pipeline}, the proposed algorithm comprises four components: phase classification, motion correction, perfusion segmentation and TICI scoring. First, given four DSA acquisitions for each patient, both anteroposterior (AP) and lateral views for both pre-EVT and post-EVT, each of these acquisitions is separated into four phases, being non-contrast, arterial, parenchymal and venous phase. We trained a multi-channel CNN model to label image phases based on spatio-temporal textural features, as well as sequence label transition rules. Subsequently, venous phase frames are removed from the sequence. Non-contrast frames are also excluded, as they contain no valid information, but merely motion artifacts. The remaining sequence frames are aligned based on affine registration to correct motion artifacts. Third, a single 2D minimum intensity map (MINIP) is calculated from the aligned sequence, which is then segmented into vessels, perfused area and non-perfused (including background) area. Finally, by comparing the perfusion area before and after EVT treatment, a quantitative TICI score is deduced as the percentage of re-perfused area out of previously occluded downstream territory. Each of the components is further detailed below.
      
\subsection{Phase Classification}\label{subsec:phase_classification}
The purpose of the phase classification step is to recognize and separate the venous phase frames; these frames need to be excluded from the series, because these may contain late perfusion attributed to retrograde flow via pial collaterals.\par

In a broad sense, phase classification can be seen as a video sequence labelling problem. Although 2D CNNs have been shown powerful in extracting representative features from images, temporal information is discarded when handling videos. Many recent studies have investigated how to effectively incorporate temporal frame dependencies in videos sequence labelling, various innovations have been proposed. Karpathy~\etal~\cite{karpathy2014large} explored feature fusion of multiple frames. Adding recurrent layers \cite{yue2015beyond, donahue2015long} have been shown effective in capturing frame ordering and long-range dependencies. Alternatively, 3D CNNs have been studied several times \cite{taylor2010convolutional, tran2015learning} for spatio-temporal feature extraction. More two-stream networks have been recently proposed to capture low-level motion based on optical flow~\cite{simonyan2014two, carreira2017quo}. A summary of related works can be found in~\cite{carreira2017quo}. Above methods focus on learning a global spatio-temporal description which best represents a video sequence. In this work, we are especially interested in detecting the phase transitions between individual frames. On one hand, temporal dependencies do provide valuable hint for labelling current frame. One the other hand, such video descriptions are not necessarily sensitive to frame shifts.\par

Specifically on phase classification, existing methods can be grouped into two types: supervised and unsupervised. Lee~\etal~\cite{lee2017automatic} used independent component analysis (ICA) to separate a DSA acquisition into three images (arterial, parenchymal and venous). Such unsupervised methods rely on the assumption of an expected number of components. For supervised methods, Schuldhaus~\etal~\cite{schuldhaus2011classification} utilized Rosenblatt perceptron-based classification using handcrafted positional intensity features. \par

\begin{figure*}[!ht]
     \centering
        \includegraphics[clip, trim=0cm 0.85cm 0cm 0.85cm, width=\textwidth]{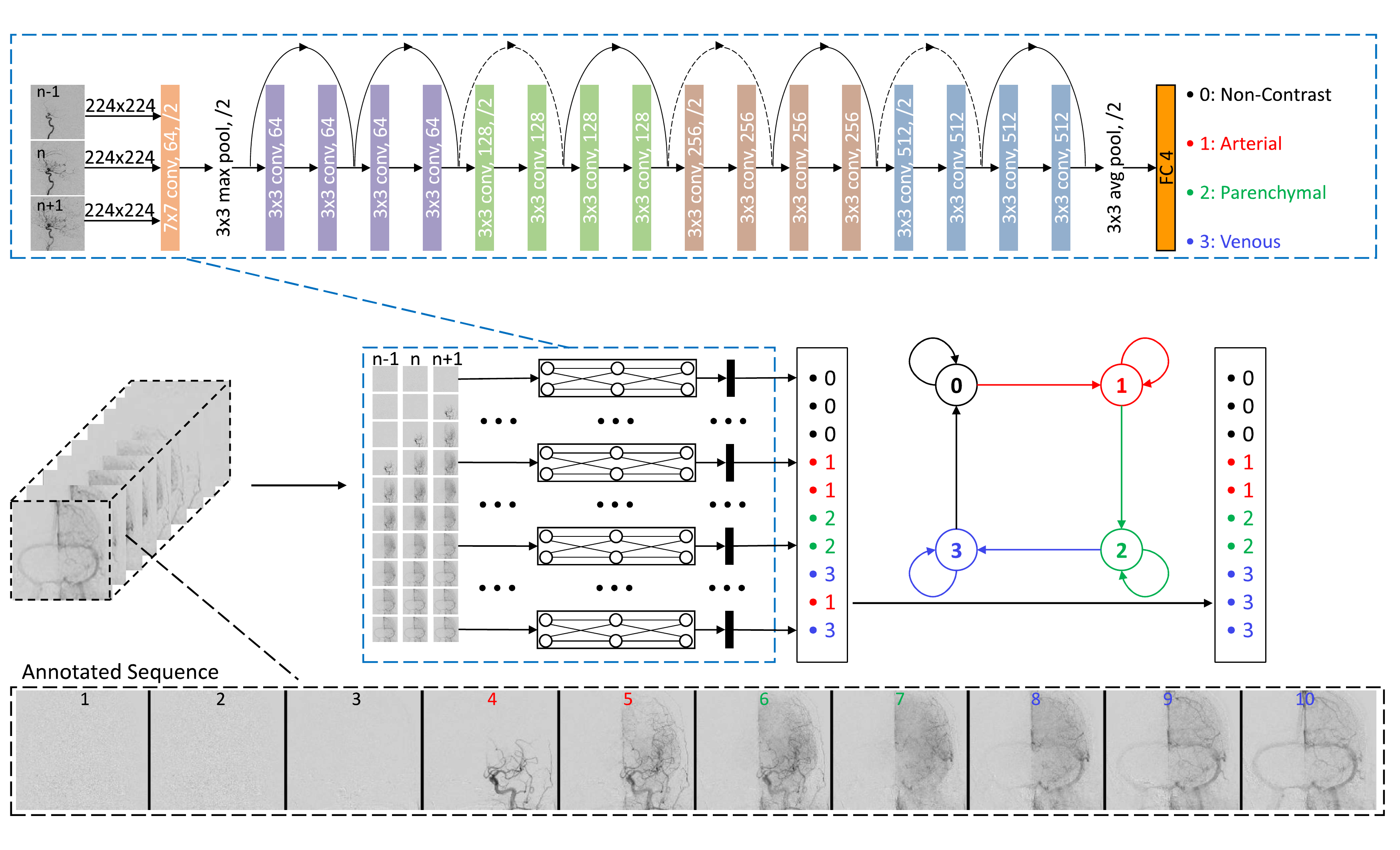}
        \caption{An overview of the phase classification framework.}
        \label{fig:phase_classification_alg}
\end{figure*}

In this work, we propose a deep learning framework using a CNN with a customized CRF for automatic phase classification. An overview of the framework is shown in Fig.~\ref{fig:phase_classification_alg}. Provided a DSA sequence, a retrained ResNet18 takes each frame together with its direct neighbors as input and predicts a phase label. Next, on the sequence level, a tailored state-transition diagram is applied on the predicted label probabilities to ensure a logical label sequence based on maximum likelihood.

\begin{figure}[!htb]
     \center{\includegraphics[trim=0cm 0cm 5cm 0cm, width=\columnwidth]
        {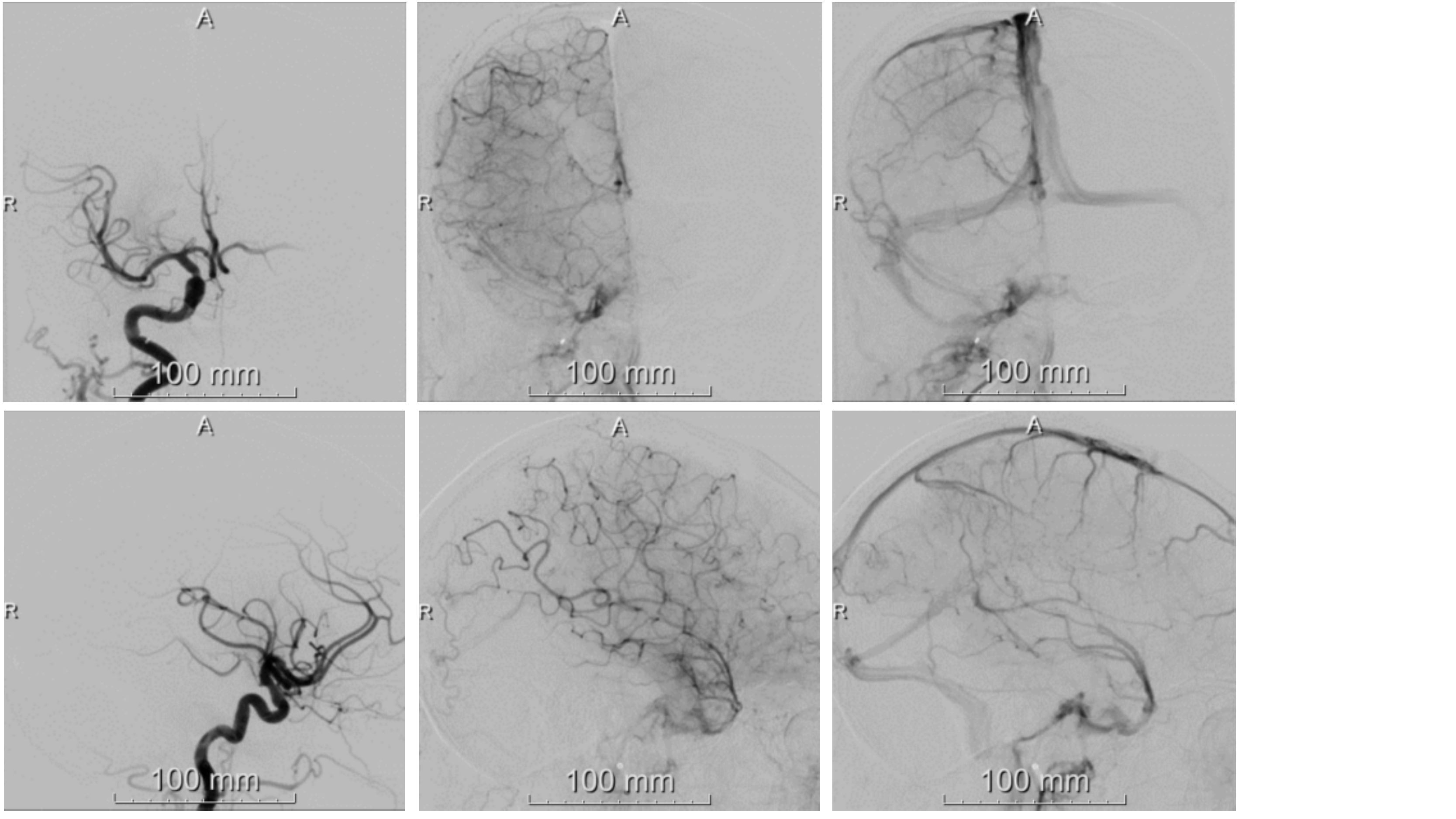}}
        \caption{DSA examples of arterial (left column), parenchymal (middle column) and venous (right column) phases. Top row: frontal view; bottom row: AP view.}
        \label{fig:phase_examples}
\end{figure}

\subsubsection{Phase Definition}\label{subsubsec:phase_definition}
A complete run of a DSA acquisition can be categorized into three phases based on the blood flow: the arterial, parenchymal and venous phase. Fig.~\ref{fig:phase_examples} shows an example of each phase in AP and lateral views. In practice, the first or last few frames generally have no visible contrast. In this work, we consider non-contrast as an additional phase. The phases were defined according to the following criteria (determined by a clinical expert):
\begin{itemize}
    \item Arterial phase:
    \begin{itemize}
        \item First frame: appearance of contrast in image;
        \item Last frame: appearance of contrast in cortical arterial branches. These are the peripherally located arterial branches which run on the cortex of the brain;
    \end{itemize}
    \item Parenchymal phase:
    \begin{itemize}
        \item First frame: right after last arterial frame.
        \item Last frame: right before appearance of contrast in the superior sagittal sinus (SSS);
    \end{itemize}
    \item Venous phase:
    \begin{itemize}
        \item First frame: appearance of contrast in the SSS;
        \item Last frame: the last frame with clearly visible contrast thereafter.
    \end{itemize}
\end{itemize}

It should be noted that a complete run of DSA does not necessarily contain all four phases. Whether these phases are all present depends on the level of the occlusion, the timing of the acquisition with respect to the contrast injection and the frame rate. In case of a carotid-T occlusion, mostly only an arterial phase is present because no contrast penetrates the affected hemisphere from the vessel where contrast was injected, and in some short acquisitions, the venous phase may not be present.
 
\subsubsection{Network Architecture}\label{subsubsec:network_architecture}
The proposed network architecture is shown in Fig.~\ref{fig:phase_classification_alg}. The network is based on ResNet-18 proposed by He~\etal~\cite{he2016deep}, to which we made modifications to adapt it to our task. In our case, the input is set to three consecutive 2D sequence frames, each for one input channel. In such a way, local features of three consecutive frames are fused. Our network ends with a fully connected (FC) layer with softmax normalization, which outputs an array of four elements with predicted per-phase probabilities.

\subsubsection{Constrained Sequence Labelling}
On the sequence level, a handcrafted state transition diagram (see Fig.~\ref{fig:phase_classification_alg}) is applied to suppress invalid phase transitions from frame to frame. The rationale is that although the network fuses neighboring frames which possess temporal contrast flow information, the global sequence level information is not fully exploited. For instance, in a correctly ordered sequence, no arterial phase should appear during or after parenchymal phase. All sequence labels should follow a certain state transition logic. We embed this logic to bring further robustness to the network. \par

In this work, we adopted the idea of conditional random fields (CRFs)~\cite{laffertyCrf} and tailored it according to our task. CRFs train a transition matrix between previous and current labels by maximum likelihood learning. Instead of training the conditional dependency between frames in a data-driven manner, we handcrafted the structural logic between labels directly into a transition matrix. The state transition diagram in Fig.~\ref{fig:phase_classification_alg} shows the allowed transitions between phases, which translates into the following transition matrix, where all allowed transitions are denoted as $T_{i,j} = 1$:

\begin{equation}
T = 
\begin{pmatrix}
1 & 1 & 0 & 0 \\
0 & 1 & 1 & 0 \\
0 & 0 & 1 & 1 \\
1 & 0 & 0 & 1
\end{pmatrix} \quad.
\label{eq:transition_matrix}
\end{equation}

We intentionally assigned equal weights to all allowed transitions ($T_{i,j}=1$) to avoid bias introduced by the training data itself. The remaining decoding and inference steps were kept the same as in CRF.\par

\begin{figure}[ht]
\centering
\begin{subfigure}[t]{.5\columnwidth}
  \centering
  \includegraphics[width=.94\linewidth]{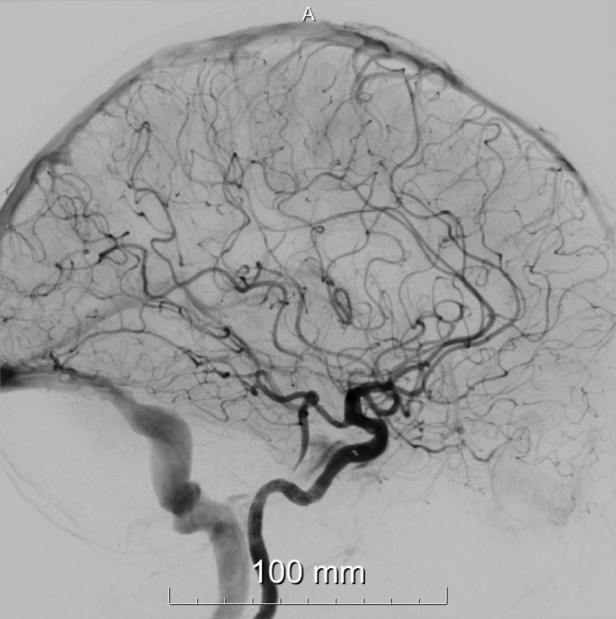}
  \caption{before.}
  \label{fig:phase_separation_example_sub1}
\end{subfigure}%
\begin{subfigure}[t]{.5\columnwidth}
  \centering
  \includegraphics[width=.95\linewidth]{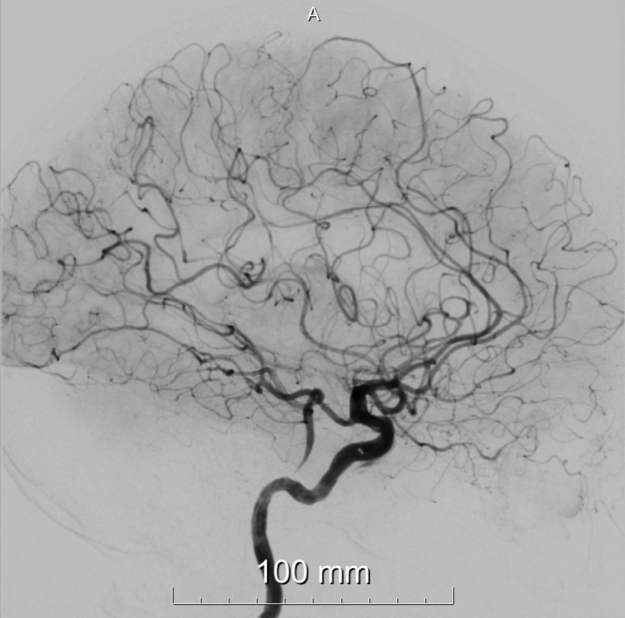}
  \caption{after.}
  \label{fig:phase_separation_example_sub2}
\end{subfigure}
\caption{A MINIP before (a) and after (b) venous phase removal.}
\label{fig:phase_separation_example}
\end{figure}

After phase classification, only arterial and parenchymal phase frames are carried on for further processing. Fig.~\ref{fig:phase_separation_example} shows an example of the venous phase removal effect.\par

\subsection{Motion Corrected MINIP}\label{subsec:motion_correction}
The image segmentation and quantification steps rely on the MINIP, which is a 2D image with each pixel being its minimum intensity value (which reflects the attenuation caused by iodine contrast agent) across the time axis of the sequence. Patient movements (voluntary or involuntary) during image acquisition may hamper this quantification, as they may cause vessels to be less apparent, and to overlap with brain tissue in the MINIP image. Therefore, the purpose of motion correction is to correct frame misalignments introduced by patient motion during image acquisition.\par

\begin{figure}[ht]
\centering
\begin{subfigure}[t]{.5\columnwidth}
  \centering
  \includegraphics[width=.95\linewidth]{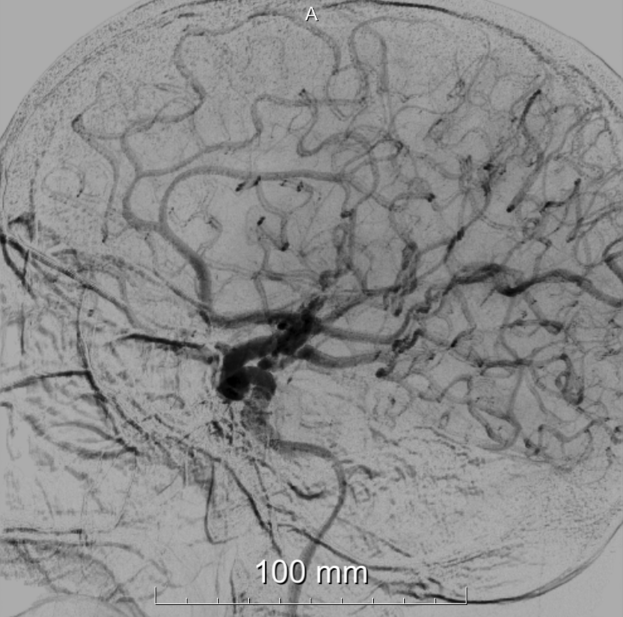}
  \caption{without motion correction.}
  \label{fig:sub1}
\end{subfigure}%
\begin{subfigure}[t]{.5\columnwidth}
  \centering
  \includegraphics[width=.95\linewidth]{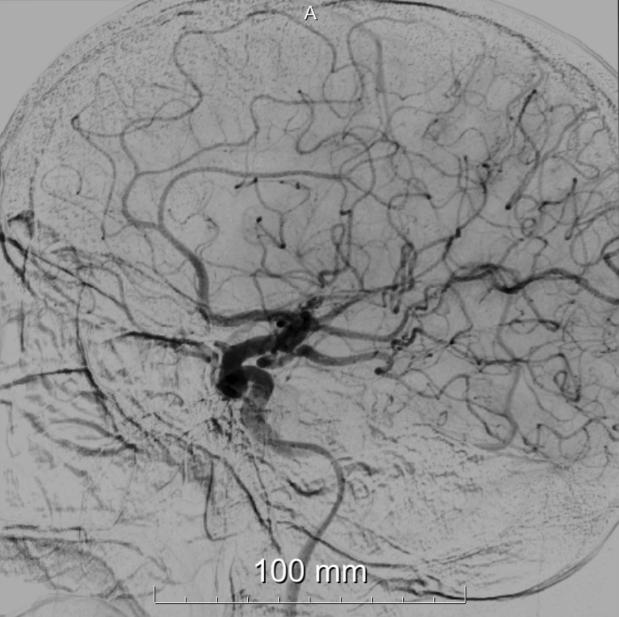}
  \caption{with motion correction.}
  \label{fig:sub2}
\end{subfigure}
\caption{A MINIP image without (a) and with (b) motion correction.}
\label{fig:elastix_example}
\end{figure}

In this work, we used an affine registration that optimizes the Mattes mutual information~\cite{mattes2003pet} with the adaptive stochastic gradient descent (ASGD)~\cite{klein2009adaptive} optimizer. Affine registration was chosen to handle possible scaling and sheer while avoiding additional artifacts that could be caused by B-spline registration. Mattes mutual information could allow us to handle contrast flow differences between frames. Motion correction was performed using the SimpleElastix toolbox~\cite{marstal2016simpleelastix}. The parameter file used in this work can be found at \url{https://elastix.lumc.nl/modelzoo/par0063/}. Fig.~\ref{fig:elastix_example} visualizes the sequence motion compensation on the calculated MINIP image.\par

\subsection{Perfusion Segmentation}\label{subsec:perfusion_segmentation}
The result of the previous steps is a MINIP of the motion corrected DSA sequence, that only contains arterial and parenchymal phases. From the constructed 2D MINIP image, (see Fig.~\ref{fig:autotici_pipeline}), we segment blood vessels and perfused area from the image background, which serves as a prerequisite for the subsequent (re-)perfusion quantification. Firstly, the contrast-opacified blood vessels are segmented using the multi-scale Frangi filter~\cite{frangi1998multiscale}, which is a common technique for enhancing vessel-like structures. In this work, the scale of Frangi vesselness filter was set to $\sigma_{min}$ = 2 and $\sigma_{max}$ = 12 according to the vessel diameter range from 5 to 35 pixels. We chose six different scales (scale step = 2) for detection of vessels of various sizes. Default value of $\alpha$ = 0.5 was used. For $\gamma$ we chose 15 based on visual inspection on a set of representative images. An intensity threshold of 0.08 was applied on filter output to segment vessel structures. We opted for such a proven vessel segmentation approach instead of deep learning methods, as we do not rely on high precision vessel segmentation, rather we exclude the vessel pixels such that the perfusion area can be better segmented. Such a traditional algorithm satisfied our needs. Next, the remaining pixels are automatically clustered into two groups, perfused and non-perfused (including background), using Otsu's thresholding technique~\cite{otsu1979threshold}, which is an effective and parameter free method based on the image intensity histogram. Image background is included in the histogram so that perfused pixels can be recognized. Finally, a segmentation colormap is constructed (see Fig.~\ref{fig:autotici_pipeline}), where vessels, perfused and non-perfused pixels are represented in red, green and blue respectively. 

\subsection{TICI Quantification}\label{subsec:tici_quantification}
\begin{figure}[htb]
\centering
\begin{subfigure}[t]{.33\columnwidth}
  \centering
  \includegraphics[width=.99\linewidth]{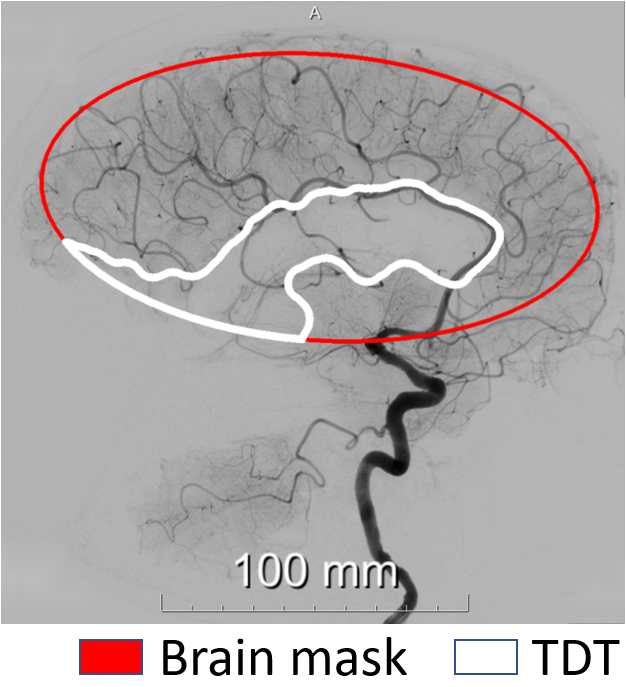}
  \caption{pre-EVT MINIP}
  \label{fig:tici_quantification_r0014_a}
\end{subfigure}%
\begin{subfigure}[t]{.33\columnwidth}
  \centering
  \includegraphics[width=.99\linewidth]{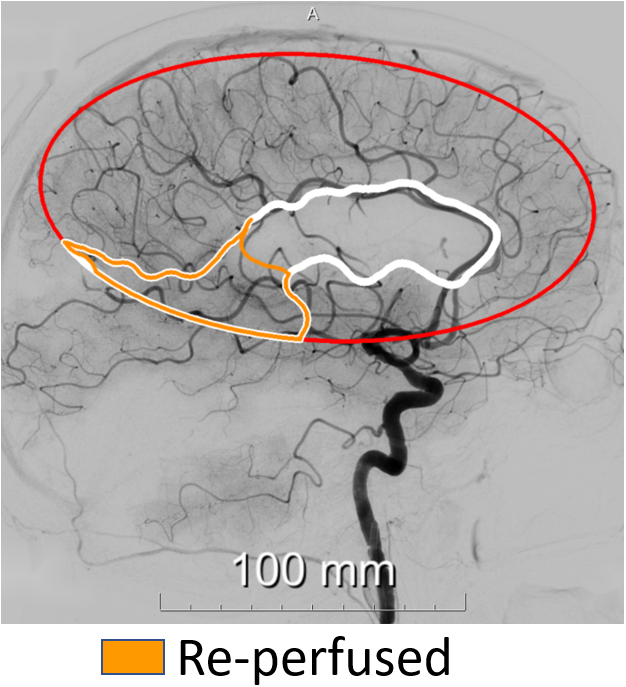}
  \caption{post-EVT MINIP}
  \label{fig:tici_quantification_r0014_b}
\end{subfigure}%
\begin{subfigure}[t]{.33\columnwidth}
  \centering
  \includegraphics[width=.99\linewidth]{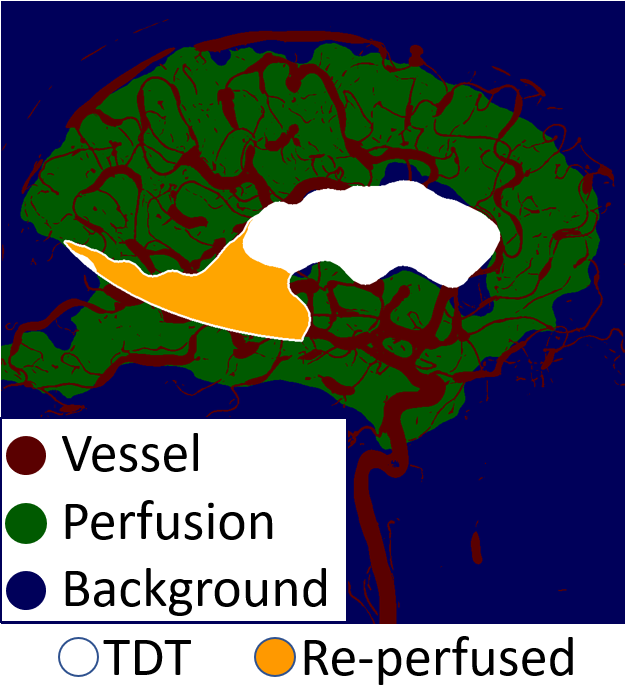}
  \caption{\centering post-EVT segmentation}
  \label{fig:tici_quantification_r0014_c}
\end{subfigure}
\caption{An example of TICI quantification. Red: brain mask; white: $TDT_{preEVT}$; orange: reperfused area ($TDT_{preEVT} \cap P_{postEVT}$).}
\label{fig:tici_quantification_r0014}
\end{figure}

Finally, a quantitative TICI score can be computed from the resulted segmentation colormap. Both AP and lateral views are evaluated. Traditional visual TICI-like scores assess the change in the level of perfusion after EVT with respect to pre-EVT. Similarly, the proposed autoTICI score quantifies the ratio of re-perfused area versus the initial TDT, which can be formulated as follows:
\begin{equation}
\mathit{autoTICI} = \frac{\mathit{TDT_{preEVT}}\cap \mathit{P_{postEVT}}}{\mathit{TDT_{preEVT}}}\quad,
\label{eq:autotici}
\end{equation}
where $\mathit{P_{postEVT}}$ denotes perfused pixels after EVT (green and orange area in Fig.~\ref{fig:tici_quantification_r0014_c}) and $\mathit{TDT_{preEVT}}$ is the initially occluded brain area prior to EVT (white area  in Fig.~\ref{fig:tici_quantification_r0014}). While $\mathit{P_{postEVT}}$ is available in the segmentation colormap, the distinction between non-perfused pixels and pure image background in the pre-EVT image is yet to be made. To this end, we introduce an atlas-based approach to exclude the out-of-skull pixels. 
More specifically, we utilize masked DSA atlases to align the brain area from atlas to the post-EVT MINIP of patients. 22 DSA acquisitions were selected as atlases from patients without stroke (Section~\ref{subsubsec:data.annotation.atlas}). We applied affine registration (similar to the registration used in motion correction in Section \ref{subsec:motion_correction}) to spatially align the atlases with the post-EVT image. The atlas brain mask was used as the moving image mask during registration. The final registered atlas for each subject is the one with the maximum Mattes mutual information~\cite{mattes2003pet}. The pre-EVT MINIP is also registered to post-EVT MINIP. In such a way, the brain region mask on the registered atlas can be mapped to pre- and post-EVT MINIP. $\mathit{TDT_{preEVT}}$ and $\mathit{P_{postEVT}}$ are determined within the aligned brain mask. As illustrated in Fig.~\ref{fig:tici_quantification_r0014}, autoTICI represents the ratio of reperfused pixels (orange area) versus the pre-EVT TDT (white area).

\section{Data and Annotation}\label{sec:data}

\subsection{Data Selection}
In this work, we used the MR CLEAN Registry~\cite{jansen2018endovascular} dataset for training and evaluating the proposed methods. The MR CLEAN Registry is an on-going multi-center registry which contains all patients with acute ischemic stroke who underwent EVT in the Netherlands since March 2014 \cite{jansen2018endovascular}. Due to the large variety in acquisition systems and imaging protocols in different centers as well as differences in patient condition, the acquired DSA images possess great variability in image appearance and quality. Therefore, it is necessary to perform data selection in order to obtain a set of qualified yet representative data for our study.
\subsubsection{Phase Classification Dataset}\label{subsubsec:data.phase_classification}
For training and evaluating the proposed phase classification algorithm, the following selection criteria were applied:
\begin{itemize}
    \item \textbf{Sequence length:}~due to non-uniformed data collection procedures across intervention centers, not all DSA acquisitions were collected in the format of a sequence; often only a few snapshots were stored. Moreover, we expect a qualified image sequence to contain most of the contrast flow. Based on our observation, acquisitions with less than six frames are insufficient for our purpose, thus excluded;
    \item \textbf{Image quality:}~for the data to be usable, a number of quality criteria must be met: (i) the acquisition must be a cerebral DSA; (ii) the acquisition must possess a sufficiently visible amount of contrast; (iii) the acquisition should not exhibit substantial motion artifacts, such as motion blur, which is not trivial to eliminate via post-processing;
    \item \textbf{Occlusion location:}~in this study, we consider patients that had an occlusion of the intracranial internal carotid artery (ICA) or the M1 segment, as these are the most common indications for EVT.
\end{itemize}

The MR CLEAN Registry (part 1) contains data of 1488 patients, which are collected from March 16, 2014 till June 15, 2016. DSA acquisitions have been stored for 1479 patients, out of which 987 patients had an occlusion of either ICA or M1. After removing all short acquisitions (less than six frames), 872 patients remained. Subsequently, 192 patients were excluded due to lack of good quality image sequences, which led to 680 qualified patients. From each of those, we selected up to four acquisitions (pre-EVT/post-EVT, AP/lateral view) per subject which were of good image quality as described above.\par

For the purpose of phase classification, we ended up with 680 qualified patients, 1857 acquisitions and 30297 images. Non-subtracted, duplicated, inverted, corrupted frames were removed and finally all acquisitions were re-ordered according to frame acquisition time.\par

\subsubsection{TICI Quantification Dataset}\label{subsubsec:data.tici_quantification}
To obtain a qualified DSA dataset for TICI quantification experiments and the overall algorithm evaluation, we applied further data selection criteria on top of the selected dataset for phase classification as follows:
\begin{itemize}
    \item \textbf{Reference eTICI score:} in this study, eTICI score was used as the reference standard for evaluating the proposed autoTICI. Therefore, it is required that both pre-EVT and post-EVT acquisitions were scored. In addition, a pre-EVT eTICI score of 2 or higher indicates that the occlusion has been reopened before EVT treatment, thus these images were excluded for further analysis;
    \item \textbf{Four acquisitions per patient:} in order to investigate the sensitivity of autoTICI on both AP and lateral views, pre-EVT and post-EVT DSA acquisitions for both views should be available;
    \item \textbf{Image quality:} for accurate TICI quantification, the acquisition should fulfill the following image quality criteria: (i) TDT of the brain should be fully visible in the image view; (ii) the view of the acquisition should not be substantially rotated. As our DSA images are 2D projections of the brain, affine transformations are unable to fix 3D rotation artifacts in its 2D projects.
\end{itemize}

Out of the 680 qualified patients from the previous selection, 34 patients were excluded due to no valid eTICI score. Following it, image quality criteria were applied and patients with four acquisitions available were selected, which leads to 141 patients (564 acquisitions) for TICI quantification and testing of phase classification. \par

\subsubsection{Dataset Division}
The statistics of selected patient information is summarized in Table~\ref{tab:dataset_distribution}. Based on data quality and availability, a subset of 141 out of 680 patients were suitable for automatic TICI quantification. This subset was fully annotated and used for testing of phase classification, as well as evaluation of the entire pipeline. To ensure good model generalizability, this test set was thus excluded from the training set for phase classification. From the remaining 539 patients, 648 acquisitions (9440 images) were randomly selected (to alleviate manual annotation effort) for training and evaluating the phase classification model. In this way, the test set is maximized and kept unseen from the trained model.\par

\begin{table}
\caption{eTICI and mRS score distribution on 141 test patients.}
\label{tab:dataset_distribution}

\begin{center}

\begin{tabular}{>{\centering\arraybackslash}m{1.2cm} R{0.3cm} R{0.8cm} ||
>{\centering\arraybackslash}m{0.2cm}|
>{\centering\arraybackslash}m{1.5cm} R{0.3cm} R{0.8cm}}

\hline
\multicolumn{7}{l}{Training set: 539 patients, 648 acquisitions, 9440 images} \\
\multicolumn{7}{l}{Test set: ~~~~~141 patients, 564 acquisitions, 7421 images} \\ \hline \hline
\multicolumn{7}{c}{Statistics on test set} \\ \hline
post-EVT eTICI & \multicolumn{2}{c||}{\specialcell{Number\\of patients}} & \multicolumn{2}{c}{mRS score} & \multicolumn{2}{c}{\specialcell{Number\\of patients}} \\ \hline
& & & \multirow{3}{*}{\rotatebox[origin=c]{90}{\parbox{1.2cm}{favorable}}} & \multirow{2}{*}{\specialcell{0~~~\\(no symptoms)}} & \multirow{2}{*}{6} & \multirow{2}{*}{(4\%)} \\
0  & 14 & (10\%)   &  &  &  &  \\
1  & 4 & (3\%)    && 1     & 20 & (14\%)   \\
2A & 24 & (17\%)   && 2     & 37 & (26\%)   \\ \cline{4-7}
2B & 27 & (19\%)   &\multirow{4}{*}{\rotatebox[origin=c]{90}{\parbox{1.5cm}{unfavorable}}} & 3     & 18 & (12\%) \\
2C & 22 & (16\%)   && 4     & 12 & (8\%)   \\
3  & 50 & (35\%)   && 5     & 6 & (4\%)    \\
  &    &&& \specialcell{6\\(death)}   &42 & (29\%)  \\

\hline
\end{tabular}
\end{center}
\end{table}

\subsection{Data Annotation}\label{subsec:data.annotation}
\subsubsection{TICI Score Reference Standard}
In the MR CLEAN registry, the eTICI~\cite{almekhlafi2014not} was visually scored by independent core-lab neuroradiologists. In this work, we compared the proposed autoTICI against eTICI.

\subsubsection{Modified Rankin Scale (mRS)}\label{subsubsec:data.annotation.mRS}
The mRS score assesses the neurological independence of patients at day 90 following an EVT. It is a seven-grade scale, running from no symptoms (score: 0) to death (score: 6). In this study, we used the mRS score as a reference to evaluate the outcome predictability of the proposed autoTICI. Table~\ref{tab:dataset_distribution} describes the mRS score distribution on the test data set. Note that 20 patients had missing mRS records, which were imputed using multiple imputations by chained equations (MICE)~\cite{azur2011multiple}. We dichotomized the mRS scores to indicate favorable and unfavorable treatment outcomes.

\subsubsection{National Institutes of Health Stroke Scale (NIHSS)}\label{subsubsec:data.annotation.NIHSS}
NIHSS quantifies the impairment of stroke patients by evaluating 11 aspects. It is an ordinal score ranging from 0 (no symtoms) to a maximum of 42. In the MR CLEAN registry, both the baseline (prior to EVT, $\mathit{NIHSS_{BL}}$) and follow-up (within 24 hours after EVT, $\mathit{NIHSS_{FU}}$) scores were assessed. In our experiments, we derived the NIHSS shift as follows:
\begin{equation}
\mathit{NIHSS_{shift}} = \mathit{NIHSS_{BL}} - \mathit{NIHSS_{FU}}\quad.
\label{eq:nihss_shift}
\end{equation}

\subsubsection{Sequence Phase Labelling}
Manual phase labelling was performed independently by three annotators, one experienced clinician, one colleague researcher and the first author, using an in-house developed tool in MevisLab. According to the phase definition (Section~\ref{subsubsec:phase_definition}), the sequence labelling output is a per frame label sequence, with 0, 1, 2, 3 denoting non-contrast, arterial, parenchymal and venous phase respectively. 648 randomly selected training sequences (Section~\ref{subsubsec:data.phase_classification}) were randomly split into 3 parts, each annotated by one operator. 141 testing sequences (Section~\ref{subsubsec:data.tici_quantification}) were labelled by all three operators independently, and consensuses were derived afterwards.\par

\subsubsection{DSA Atlas}\label{subsubsec:data.annotation.atlas}
We selected 22 DSA acquisitions (12 AP and 10 lateral views) from 50 patients without stroke. Binary brain masks were delineated on those atlases for registration purposes as described in Section~\ref{subsubsec:data.tici_quantification}.

\section{Experiments and Results} \label{sec:exp&result}

\subsection{Implementation}
The proposed methods were implemented in Python. The deep learning model for phase classification was developed using PyTorch~\cite{paszke2019pytorch} on an NVIDIA 2080 Ti with 11 GB of memory. The deep learning network was trained with a batch size of 32 for 100 epochs. All training data were iterated once per epoch. To enrich the diversity of training set and prevent possible overfitting, the following augmentation techniques were randomly applied during data loading: horizontal flip, random rotation ($\in [-10^{\circ}, 10^{\circ}]$), random affine transformation (translation $\in [0, 10\%]$ of image width/height, scale $\in [0.8, 1.2]$). For this multi-class classification task, cross entropy was chosen as the loss function. We used the Adam optimizer~\cite{kingma2014adam} with an adaptive learning rate initialized at 0.001, halved every 10 epochs.\par

\subsection{Evaluating Phase Classification}
In the evaluation of phase classification, we assessed the added value of data augmentation, temporal information and the handcrafted constrained transition matrix module via an ablation study. Furthermore, we compared the method with inter-observer variability among three human annotators. Additionally, the impact of automatic phase classification accuracy on autoTICI scoring is presented in \ref{subsubsec:phase_classification_impact_on_autoTICI}. 

\begin{table*}[ht]
\begin{center}
\caption{Performance overview of phase classification. Ablation study was performed with 5-fold cross validation on the training and validation set. Performance comparison with human annotator was assessed on the test set.}
\label{tab:phase_classification_results}

\begin{tabular}{|l|l|c|c|m{0.3cm}m{0.95cm}m{1.05cm}|m{0.3cm}m{0.95cm}m{1.05cm}|m{0.3cm}m{0.95cm}m{1.05cm}|}
\hline
\multicolumn{2}{|c|}{\multirow{3}{*}{Method}}
& \multicolumn{2}{c|}{Frame level metrics}
& \multicolumn{9}{c|}{Sequence level metrics} \\ \cline{3-13} \cline{3-13}

\multicolumn{2}{|c|}{} & \multirow{3}{*}{\specialcell{Average\\accuracy}} 
& \multirow{3}{*}{\specialcell{Weighted\\F1}}
& \multicolumn{3}{c|}{first arterial} & \multicolumn{3}{c|}{last arterial} & \multicolumn{3}{c|}{last parenchymal} \\ \cline{5-13}

\multicolumn{2}{|c|}{}& & & \multirow{2}{*}{acc} & \multicolumn{2}{c|}{offset: frame(sec\textsuperscript{*})} & \multirow{2}{*}{acc} & \multicolumn{2}{c|}{offset: frame(sec\textsuperscript{*})} & \multirow{2}{*}{acc} & \multicolumn{2}{c|}{offset: frames(sec\textsuperscript{*})} \\

\multicolumn{2}{|c|}{}& & & & \multicolumn{1}{c}{mean} & \multicolumn{1}{c|}{std} & & \multicolumn{1}{c}{mean} & \multicolumn{1}{c|}{std} & & \multicolumn{1}{c}{mean} & \multicolumn{1}{c|}{std} \\ \hline

\multirow{4}{*}{\rotatebox[origin=c]{90}{Ablation}} & Proposed   & \textbf{0.92}  & \textbf{0.92} & \textbf{0.92} & \textbf{0.10}(0.05)  & \textbf{0.41}(0.16) & \textbf{0.51}  & \textbf{0.70}(0.53) & \textbf{1.20}(0.94) & \textbf{0.66}  & \textbf{0.42}(0.39) & \textbf{0.84}(1.15)  \\ \cline{2-13}

& No augmentation & 0.88  & 0.88  & 0.84 & 0.21(0.09)  & 0.67(0.25) & 0.39 & 1.00(0.81) & 1.56(1.65) & 0.60  & 0.53(0.54) & 0.98(1.32)  \\ 

& No temporal info & 0.91  & 0.91 & 0.91 & 0.12(0.06)  & 0.47(0.17) & 0.46  & 0.76(0.54) & 1.23(0.96) & 0.66  & 0.43(0.44) & 0.84(1.04)  \\ 

& No state-transition matrix & 0.91  & 0.91 & 0.91 & 0.14(0.06) & 0.55(0.18)  & 0.51 & 0.89(0.55) & 3.21(1.12)  & 0.66 & 0.49(0.42)  & 1.41(1.26)  \\ \hline \hline

\multirow{4}{*}{\rotatebox[origin=c]{90}{\parbox{0.5cm}{Test}}} & \multirow{2}{*}{Method to human agreement} & 0.90 & 0.90 & 0.88  & 0.14(0.05) & 0.48(0.14) & 0.34  & 1.10(0.77) & 1.54(1.56)  & 0.55 & 0.63(0.77) & 1.11(1.31)  \\ \cline{3-13}

 &  & \multicolumn{3}{r}{$>$1 frame(sec)} & \multicolumn{2}{l}{1.4\%(0.4\%)} & \multicolumn{3}{c}{24.8\%(21.3\%)}  & \multicolumn{3}{c|}{10.9\%(23.2\%)}  \\ \cline{2-13}

& \multirow{2}{*}{Human to human agreement} & 0.89 & 0.89 & 0.87  & 0.18(0.06) & 0.59(0.17) & 0.31  & 1.16(0.80) & 1.56(1.58) & 0.52 & 0.68(0.89) & 1.10(1.47)  \\ \cline{3-13}

 &  & \multicolumn{3}{r}{$>$1 frame(sec)} & \multicolumn{2}{l}{1.9\%(1.1\%)} & \multicolumn{3}{c}{25.8\%(21.0\%)}  & \multicolumn{3}{c|}{12.3\%(31.1\%)}  \\
\hline
\multicolumn{13}{l}{\footnotesize{\specialcell{\parbox{17.5cm}{\textsuperscript{*}As the frame time increments of 103 (out of 648) DSA acquisitions are missing, the time offsets are the averages of the ones with frame time increments.\\\textsuperscript{*}The frame rate is inconstant in a DSA acquisition, it typically starts at 2-4 fps in arterial phase and gradually reduces to 0.5-1 fps in venous phase.}}}}

\end{tabular}
\end{center}
\end{table*}

\subsubsection{Evaluation metrics}
The following metrics were adopted in this experiment:
\begin{itemize}
    \item \textbf{average accuracy} represents the percentage of correctly classified images out of all the images;
    \item \textbf{weighted F1 score} refers to the harmonic mean of the precision and recall. In a multi-class problem of $\mathit{Nc}$ classes, the weighted F1 ($\mathit{F1_{w}}$) is an average of F1 scores ($\mathit{F1_{i}}$) weighted by the number of images ($\mathit{S_i}$) from each class, as in Eq.~\ref{eq:weighted_f1}:
    \begin{equation}
        \mathit{F1_{w}} = \frac{\sum\limits_{i=1}^{\mathit{Nc}} \mathit{S_i F1_i}}{\sum\limits_{i=1}^{\mathit{Nc}} \mathit{S_i}}\quad;
        \label{eq:weighted_f1}
    \end{equation}
    \item \textbf{frame offset} measures the phase classification accuracy on sequence level, rather than image level. The phase border frame index offset between classified and ground truth can serve as an insightful metric for phase classification accuracy. In this experiment, the average absolute offset and standard deviation of offset for first, last arterial phase frame and last parenchymal phase frame were evaluated.
\end{itemize}

\subsubsection{Ablation Study}
Ablation experiments were performed to assess the contribution of separate components of the proposed algorithm. Table~\ref{tab:phase_classification_results} summarizes the comparison under the aforementioned metrics. In all ablated cases, the model performance degraded. Without data augmentation, the generalizability of the model was reduced as a result of overfitting on the training data. Without incorporating neighboring frames as temporal information, the overall image level accuracy slightly dropped. The most marginal deterioration was observed at arterial/parenchymal border frame accuracy (51\% vs 46\%). Not surprisingly, by embedding human defined logic into the model, an overall performance improvement was evidenced across all metrics in the Table. More importantly, such a constrained transform matrix enforces phase transition logic within a sequence, suppressing unreasonable prediction errors.\par

\subsubsection{Inter-observer Variability}
Smooth transition between phases is often observed, which attributes to inter-observer variability during phase annotation. In this section, the algorithm variability is compared with three human annotators on the selected 141 acquisitions in test set. We computed the agreements in pairs for all combinations among the proposed method and three human annotators. For the agreement between the method and human annotators is defined as the average agreement between the method and each of the annotators. The inter-observer agreement is calculated as the average agreement among all human annotator pairs. As shown in Table~\ref{tab:phase_classification_results}, the proposed method exhibits even better agreement than the inter-observer agreement across all metrics, achieving human level annotation precision on phase classification.\par

\subsection{Evaluating autoTICI Quantification}
On the test set (Table~\ref{tab:dataset_distribution}), the relevance of autoTICI as a brain tissue reperfusion measure was evaluated based on the correlation with eTICI. Furthermore, we assessed the clinical value of autoTICI in comparison to eTICI with respect to mRS and NIHSS. Whereas the distribution of patients with respect to eTICI and mRS is imbalanced, we did not attempt to balance the distribution during data selection as this represents realistic patient statistics in clinical practice.\par

\begin{figure}[!b]
\centering
\begin{subfigure}[t]{\columnwidth}
   \includegraphics[clip, trim=4.8cm 0.5cm 4.5cm 0.7cm, width=1\linewidth]{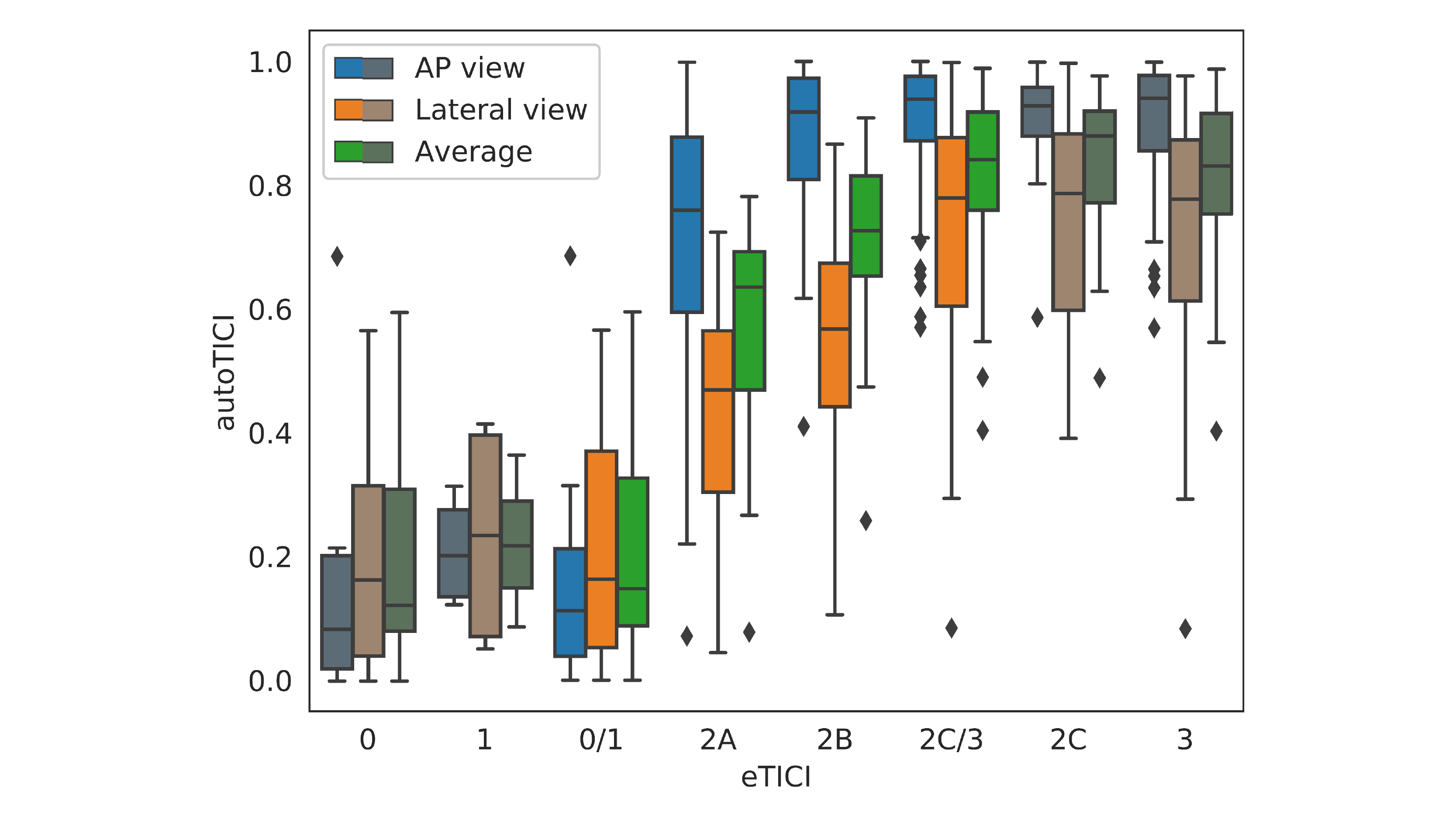}
   \caption{autoTICI distribution with respect to eTICI. Blue: AP score; orange: lateral score; green: average of AP and lateral scores.}
   \label{fig:autotici_etici_box} 
\end{subfigure}
\begin{subfigure}[t]{\columnwidth}
   \includegraphics[clip, trim=0.45cm 0.4cm 1.1cm 0cm, width=1\linewidth]{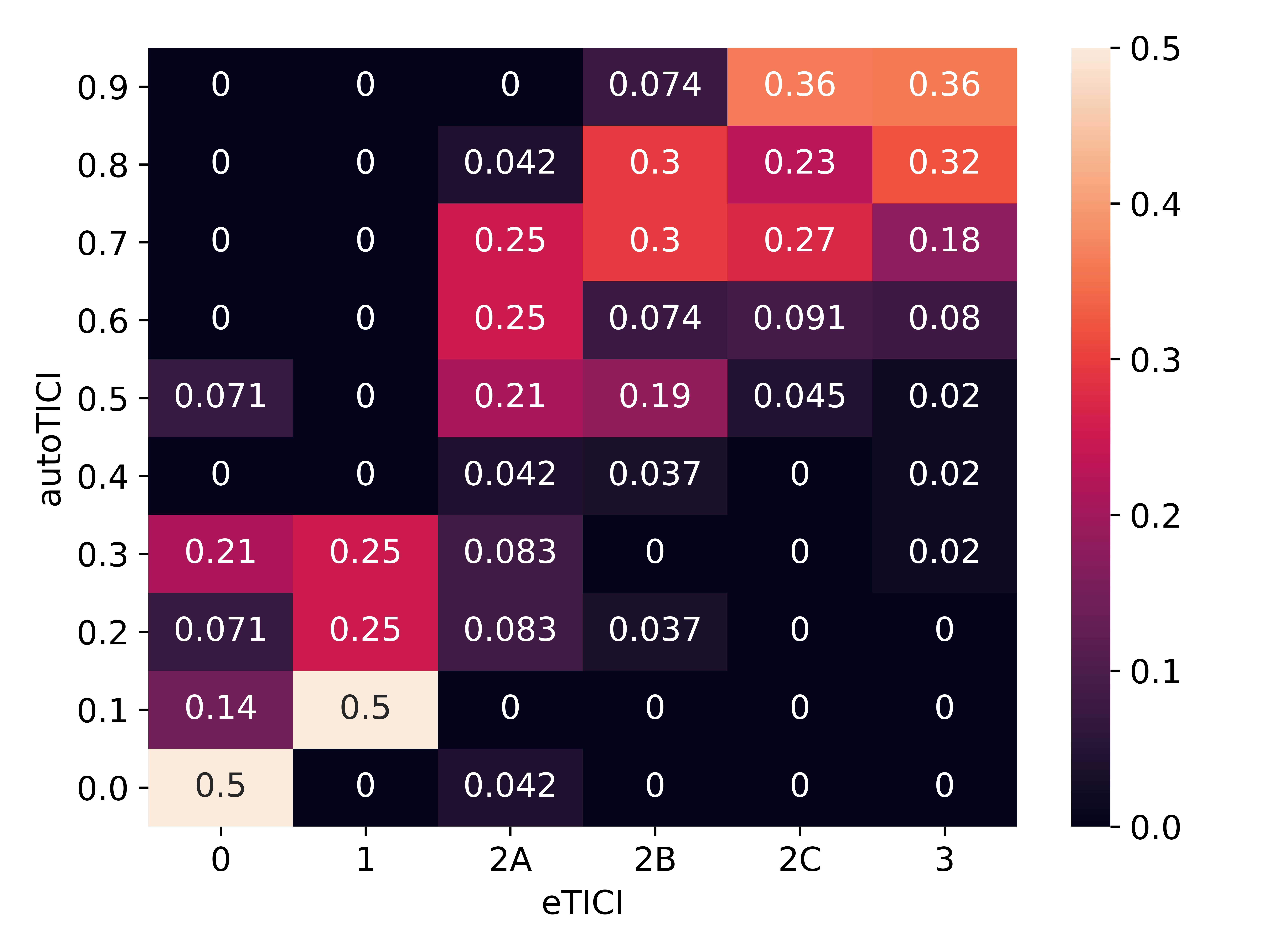}
   \caption{Distribution heatmap of autoTICI over eTICI. Here, autoTICI (y-axis) refers to the average of AP and lateral scores binned in 0.1 intervals. The value in each cell is normalized per column.}
   \label{fig:autotici_etici_heatmap} 
\end{subfigure}
\end{figure}
\begin{figure}\ContinuedFloat
\begin{subfigure}[t]{\columnwidth}
   \includegraphics[clip, trim=0.4cm 0.4cm 0.5cm 0cm, width=1\linewidth]{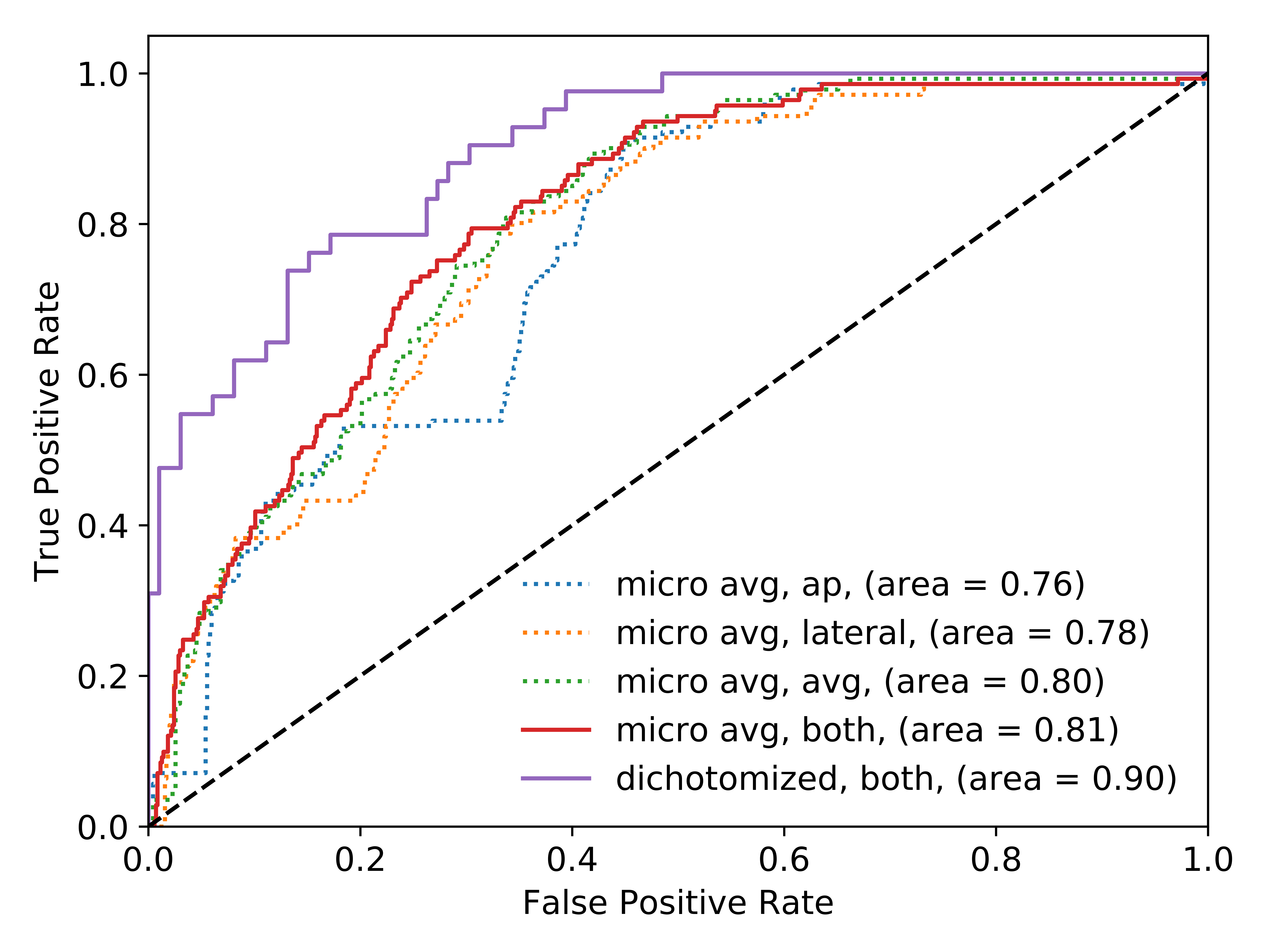}
   \caption{Logistic regression ROC. Green, orange, blue, red: micro-average ROC of all eTICI grades; purple: dichotomized eTICI (failure: $\leq 2A$; success: $\geq 2B$).}
   \label{fig:autotici_etici_log}
\end{subfigure}
\caption{Correlation between autoTICI and eTICI.}
\label{fig:autotici_etici}
\end{figure}

\subsubsection{autoTICI vs eTICI}\label{subsubsec:autoTIIC_vs_eTICI}
Fig.~\ref{fig:autotici_etici_box} shows the positive correlation between the proposed autoTICI and eTICI on the test dataset. According to eTICI definition, minimal tissue perfusion difference is expected between eTICI 0 and 1 or between eTICI 2C and 3. Therefore, eTICI 0/1 and eTICI 2C/3 are plotted to better visualize the correlation in brain tissue reperfusion. More quantitatively, Table \ref{tab:autotici_vs_etici} shows the reperfusion statistics of autoTICI versus eTICI grades. The heatmap in Fig.~\ref{fig:autotici_etici_heatmap} illustrates the distribution of autoTICI over eTICI grades. Furthermore, spearman correlation test also showed that both AP and lateral view autoTICI scores were significantly associated with eTICI with $\rho$ = 0.54, (P $<$ 0.001) and $\rho$ = 0.65, (P $<$ 0.001), respectively. The stronger correlation with lateral view autoTICI scores matches the expectation of clinical experts because the brain area in AP views is about half as in lateral projections and occlusion effects are generally better visualized in lateral view. Reasons for the few outliers shown in Fig.~\ref{fig:autotici_etici_box} and \ref{fig:autotici_etici_heatmap} include questionable eTICI annotation (recanalization without reperfusion), insufficient contrast and suboptimal TDT registration.\par

\begin{table}[t]
\centering
\caption{Quantitative autoTICI reperfusion statistics.}
\label{tab:autotici_vs_etici}
\begin{tabular}{|c|c|c|c|c|}
\hline
\multirow{2}{*}{eTICI} & \multirow{2}{*}{\specialcell{Mean area (pixels)\\TDT\_preEVT}} & \multirow{2}{*}{\specialcell{Mean area (pixels)\\TDT\_postEVT}}
& \multicolumn{2}{c|}{Reperfusion ratio} \\ \cline{4-5}

 & & & mean & std \\ \hline
 0  & \num{1.43e5} & \num{1.24e5} & 0.19 & 0.23 \\ 
 1  & \num{1.30e5} & \num{0.92e5}  & 0.22 & 0.14 \\ 
 2A & \num{1.49e5} & \num{0.70e5}  & 0.57 & 0.25 \\ 
 2B & \num{1.72e5} & \num{0.64e5}  & 0.71 & 0.23 \\
 2C & \num{1.31e5} & \num{0.26e5}  & 0.83 & 0.16 \\
 3  & \num{1.29e5} & \num{0.28e5}  & 0.82 & 0.18 \\ \hline
 \multicolumn{5}{l}{\footnotesize{\specialcell{\parbox{8.5cm}{TDT\_preEVT: non-perfused pixels before EVT; TDT\_postEVT: non-reperfused pixels after EVT; reperfusion ratio is based on Eq.~\ref{eq:autotici}}}}}
\end{tabular}
\end{table}

We performed multinomial logistic regression between autoTICI and eTICI on the test data with leave-one-out cross validation. As shown in Fig.\ref{fig:autotici_etici_log}, a micro-average (biased average by class frequency) AUC of 0.76 and 0.78 for AP and lateral view were achieved respectively. With both autoTICI scores as input features, the micro-average AUC reached 0.81. For dichotomized eTICI (failure: $\leq$ 2A; success: $\geq$ 2B), autoTICI achieved an AUC of 0.90.\par

\subsubsection{Correlation to Treatment Outcome} \label{subsubsec:outcome_prediction}
\begin{figure}[ht]
    \center{\includegraphics[clip, trim=0.4cm 0.4cm 0.5cm 0.3cm, width=\columnwidth]{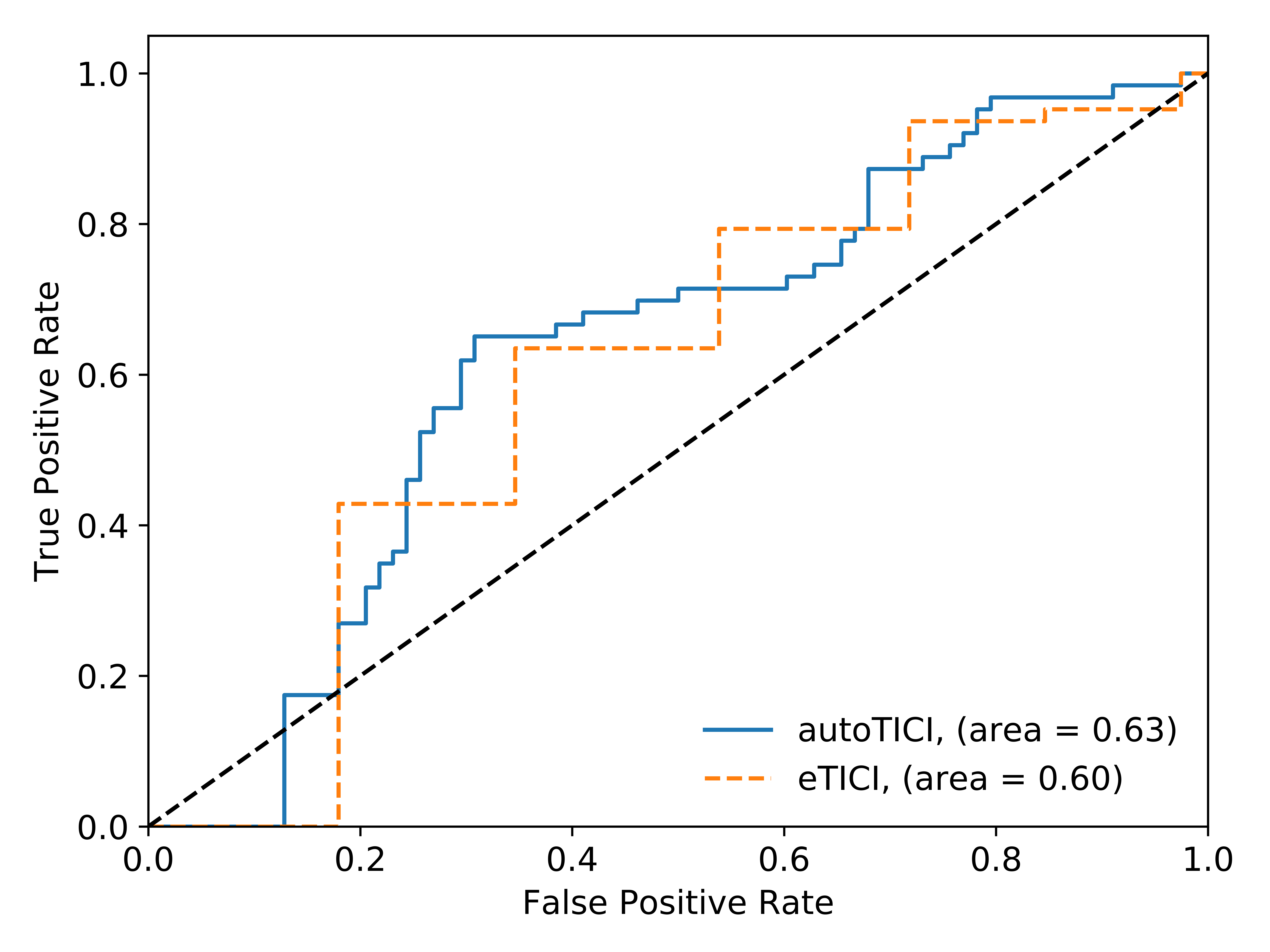}}
    \caption{Prediction of treatment outcome (mRS) using autoTICI.}
    \label{fig:autotici_mrs}
\end{figure}

We further compared autoTICI and eTICI with respect to their correlation to treatment outcome in terms of mRS (\ref{subsubsec:data.annotation.mRS}) and NIHSS (\ref{subsubsec:data.annotation.NIHSS}). We used both AP and lateral autoTICI scores as input features for logistic regression.
As shown in Fig.~\ref{fig:autotici_mrs}, cross validated logistic regression between TICI scores and mRS showed that autoTICI had a slightly higher AUC of 0.63 than 0.60 for eTICI, though the difference was not statistically significant (P = 0.52). Similarly, the accuracy of autoTICI and eTICI were 0.66 and 0.62, respectively. Spearman correlation test was applied between TICI scores and NIHSS shift scores, which showed that autoTICI and eTICI were comparable with $\rho$ = 0.29~(P $<$ 0.01) and $\rho$ = 0.30~(P $<$ 0.01), respectively.\par

\begin{table}[htb]
\centering
\caption{autoTICI vs eTICI in terms of correlation to outcome.}
\label{tab:correlation_to_outcome}
\begin{tabular}{|C{1.05cm}|C{0.5cm}|C{0.6cm}|C{0.6cm}|C{0.6cm}|C{0.75cm}|C{0.5cm}|C{0.75cm}|}
\hline
\multirow{3}{*}{\textbf{Method}}   & \multicolumn{3}{c|}{\multirow{2}{*}{\specialcell{logistic regression\\\textbf{dichotomized mRS}}}} & \multicolumn{4}{c|}{spearman correlation} \\ \cline{5-8}
&\multicolumn3{c|}{}&\multicolumn{2}{c|}{\textbf{NIHSS\_{FU}}} & \multicolumn{2}{c|}{\textbf{NIHSS\_{shift}}} \\ \cline{2-8}

& acc & AUC & odds ratio & {$\rho$} & p-value & {$\rho$} & p-value \\ \hline
autoTICI    & 0.66  & 0.63 & 1.27 & -0.25 & $<$0.01 & 0.29 & $<$0.01   \\
eTICI    & 0.62  & 0.60 & 1.02 & -0.28 & $<$0.01 & 0.30 & $<$0.01   \\ \hline
\end{tabular}
\end{table}

Overall, both Fig.~\ref{fig:autotici_mrs} and Table~\ref{tab:correlation_to_outcome} have demonstrated that autoTICI and eTICI possess comparable but limited capability in outcome prediction. It is also understandable that the relation between NIHSS (short term, at 24 hours after EVT) and mRS (long term, at 90 days after EVT) scores is not deterministic. Multiple recent studies~\cite{berkhemer2015randomized,zaidat2018primary, jansen2018endovascular} have shown that reperfusion scores play an irreplaceable role in outcome prediction, the long-term functional outcome is, however, multifactorial.\par

\subsubsection{Impact of Phase Classification on autoTICI}\label{subsubsec:phase_classification_impact_on_autoTICI}
As the correlation between autoTICI and eTICI has been established in Section~\ref{subsubsec:autoTIIC_vs_eTICI}, this correlation can be further employed to assess the value of accurate phase classification. We introduced offsets to the phase classification results and observed how autoTICI is impacted in terms of its correlation with eTICI. As shown in Table~\ref{tab:impact_of_phase_classification_acc_on_autoTICI}, adding offsets of one or two frames in classification results lead to noticeable deterioration in autoTICI statistics.\par

\begin{table}[htb]
\centering
\caption{Correlation between autoTICI and eTICI correlation with respect to phase classification accuracy}
\label{tab:impact_of_phase_classification_acc_on_autoTICI}
\begin{tabular}{|c|c|c|c|}
\hline
\multirow{2}{*}{Offset} & \multirow{2}{*}{\specialcell{Pearson correlation\\AP/Lateral}} & \multicolumn{2}{c|}{Logistic regression AUC} \\ \cline{3-4}

& & \specialcell{Micro-average\\both views} & \specialcell{Dichotomized\\eTICI} \\ \hline
PC  & 0.54/0.65 & 0.81 & 0.90 \\ 
PC$\pm$1 frame~  & 0.54/0.59 & 0.79  & 0.88 \\ 
PC$\pm$2 frames  & 0.46/0.51 & 0.76  & 0.82 \\ 
\hline
\multicolumn{4}{l}{\footnotesize{\specialcell{\parbox{8.5cm}{PC refers to the predicted results of the phase classification method.}}}}
\end{tabular}
\end{table}

\section{Discussion}\label{sec:discussion}
We have presented a fully automatic, quantitative and truly perfusion based TICI scoring method that is inspired by traditional visual TICI assessment procedures. We exploited convolutional neural networks to tackle the phase classification challenge; it achieved performance on par with human experts. On the MR CLEAN registry, we demonstrated that the proposed autoTICI and eTICI are statistically significantly associated and they are both comparable predictors of functional outcome, revealing the potential value of autoTICI as a computer-aided biomarker for peri-procedural treatment assessment.\par

We have presented a deep learning based approach for DSA phase classification with human level precision. Apart from autoTICI, the proposed phase classification method can be generalized for many automated DSA analysis tasks, such as key frames selection, inter-acquisition temporal synchronization or slow arterial flow detection. The proposed network uses ResNet with a customized CRF. Although recurrent neural networks, such as LSTM~\cite{hochreiter1997long} or GRU~\cite{cho2014learning} can be alternatives in this task, we found that the proposed CRF handles inter-frame relations well in an explainable way, especially with limited number of labelled sequences (648 variable-length sequences in our case). When considering prior-art on phase classification, only Schuldhaus~\etal~\cite{schuldhaus2011classification} reported a quantitative performance results. On 14 DSA acquisitions, Schuldhaus~\etal~\cite{schuldhaus2011classification} reported 93\% and 50\% accuracy on first arterial border match and last arterial border match respectively. Accordingly, our proposed method achieved 92\% and 51\% respectively over 648 acquisitions. A pure numerical comparison seems to show comparable performance, but our study has different characteristics. It was tested on 648 instead of 14 acquisitions, these acquisitions originate from a multi-center registry, and the frame rates (0.5-4 fps) are higher than those (0.5-2 fps) reported by Schuldhaus~\etal~\cite{schuldhaus2011classification}, and our approach also separates the venous phase.\par

The MR CLEAN Registry is an observational multi-center multi-year registry. This registry reflects daily practice in a wide variety of hospitals which on the one hand leads to the heterogeneity of data but on the other hand allows for broad translatability. It can be noticed that 141 out of 1488 patients end up suitable for autoTICI quantification. This is due to a large heterogeneity on image quality, annotation, acquisition process, and storage format as described in the data selection procedure, rather than algorithmic limitations. Main reasons for data exclusion in this study include missing DSA acquisitions, corrupted, incomplete, or too short acquisitions, bad image quality, occlusion location and image not showing the entire TDT area.\par

Affine registration was performed for sequence motion correction and atlas registration. In the MR CLEAN Registry, one of the limitations is that the original unsubtracted images were mostly discarded, leaving only the subtracted ones. In this case, motion correction does help in mitigating blur effects, it is nevertheless incapable of handling the subtraction artefacts. Besides, due to lacking of background texture, the atlas to patient registration relies on the skull outline and brain vasculature. The skull skeleton shown in a subtracted image is in fact the subtraction artifacts. Therefore, if the original non-subtracted images could be used instead, which encompass richer and more robust textural information, the registration robustness and accuracy could be potentially improved. \par

Aside from Otsu's thresholding, several alternatives for perfusion segmentation could be considered. One could quantify changes in vessels and tissue perfusion by directly subtracting the MINIP of parenchymal phase by the MINIP of arterial phase. This is however problematic, mainly due to inter-acquisition variations on contrast dilution and inject volume. It remains an interesting research topic on how to obtain contrast profile independent DSA perfusion parameters. The inter-acquisition variations could be alleviated if mechanical pumps are utilized during contrast injection with a fixed injection protocol, which however is not the case for MR CLEAN registry. Another alternative is to use the background statistics to derive a fixed threshold for TDT selection. We opted for image specific Otsu’s thresholding instead of fixed thresholding for all patients. Other promising alternatives include deep learning based segmentation methods, such as U-Net~\cite{ronneberger2015u}, which is not studied in this work due to the massive amount of annotations required.\par

eTICI is one of the most comparable metrics to autoTICI as both are proposed as brain perfusion measures. However, it should be pointed out that the definition of autoTICI and eTICI are not fully identical. While autoTICI and eTICI focus on brain tissue antegrade reperfusion quantification, eTICI also considers evidence of contrast material penetration in vessels past initial occlusion and slow flow in distal. Both eTICI 0 and 1 define minimal reperfusion, they are distinguished based on whether contrast material has passed the initial occlusion. Analogically, in case of (nearly) complete perfusion, eTICI emphasizes the existence of slow flow in distal vessels in score 2C versus 3. Therefore, perfect correlation between autoTICI and eTICI cannot be achieved and is not the aim of autoTICI. In this work, a good correlation between the two (Fig.~\ref{fig:autotici_etici}) serves as a proxy for demonstrating the relevance of autoTICI. The outcome prediction capability of autoTICI reported in Section~\ref{subsubsec:outcome_prediction} further consolidates the conclusion and reveals the potential clinical value of autoTICI.\par

End-to-end deep learning based methods might be valid alternatives to the automatic TICI scoring method proposed in this study. The reasons that we opted for a step-by-step strategy are (i) data: end-to-end training approaches treat all four acquisitions of each patient as one input data sample. Such methods generally require larger number of patients, as well as more sophisticated GPU resources; (ii) interpretability: while aiming for minimizing prediction errors, end-to-end training approaches generally sacrifices causal interpretability. The proposed method, by contrast, provides intermediate visualizable outputs per step, offering enriched clinical insights; (iii) quantitativity: rather than categorical TICI grading, we seek quantitative brain reperfusion analysis methods. Training end-to-end networks for this purpose is not straightforward due to lacking of quantitative ground truth. Nevertheless, both phase classification and perfusion segmentation can benefit from deep learning methods in this work.\par

From a clinical perspective, the proposed autoTICI provides an objective, reproducible and quantitative measure of EVT quality, eliminating human errors and variations. As it is a true measure of tissue level reperfusion, human conceptual confusion is avoided. Therefore, autoTICI overcomes the aforementioned three shortcomings of existing TICI scores. Moreover, the extent and location of non-reperfused areas are real-time visualized, helping the operator to determine whether additional attempts for clot removal should be undertaken or additional drugs should be given to improve functional outcome.\par

\section{Conclusion} \label{sec:conclusion}
We have presented a robust and fully automatic perfusion quantification method, autoTICI. On a large routinely acquired multi-center dataset, we have demonstrated that autoTICI is significantly correlated with the eTICI reference with a dichotomized AUC of 0.90 and possesses comparable treatment outcome predictive capability with an AUC of 0.63 versus 0.60, revealing its potential in future studies and clinical practice.\par

\section*{Acknowledgment}
The authors would like to thank Kars C.J. Compagne for his assistance in patient data imputation.

\bibliographystyle{IEEEtran}
\bibliography{IEEEabrv,IEEEexample}

\end{document}